\documentclass[PRB,twocolumn,groupedaddress,nofootinbib,superscriptaddress]{revtex4-1}
\usepackage{graphicx}
\usepackage{dcolumn}
\usepackage{bm}
\usepackage{times}
\usepackage[version=3]{mhchem} 

\begin{document}

\title{High temperature magneto-structural transition in van der Waals-layered $\alpha$-\ce{MoCl3}}

\author{Michael A. McGuire}
\email{McGuireMA@ornl.gov \\  \\ Notice: This manuscript has been authored by UT-Battelle, LLC under Contract No. DE-AC05-00OR22725 with the U.S. Department of Energy. The United States Government retains and the publisher, by accepting the article for publication, acknowledges that the United States Government retains a non-exclusive, paid-up, irrevocable, world-wide license to publish or reproduce the published form of this manuscript, or allow others to do so, for United States Government purposes. The Department of Energy will provide public access to these results of federally sponsored research in accordance with the DOE Public Access Plan (http://energy.gov/downloads/doe-public-access-plan). }
\affiliation{Materials Science and Technology Division, Oak Ridge National Laboratory, Oak Ridge, Tennessee 37831 USA}
\author{Jiaqiang Yan} \affiliation{Materials Science and Technology Division, Oak Ridge National Laboratory, Oak Ridge, Tennessee 37831 USA}
\author{Paula Lampen-Kelley} \affiliation{Materials Science and Technology Division, Oak Ridge National Laboratory, Oak Ridge, Tennessee 37831 USA}
\author{Andrew F. May} \affiliation{Materials Science and Technology Division, Oak Ridge National Laboratory, Oak Ridge, Tennessee 37831 USA}
\author{Valentino R. Cooper} \affiliation{Materials Science and Technology Division, Oak Ridge National Laboratory, Oak Ridge, Tennessee 37831 USA}
\author{Lucas Lindsay} \affiliation{Materials Science and Technology Division, Oak Ridge National Laboratory, Oak Ridge, Tennessee 37831 USA}
\author{Alexander Puretzky} \affiliation{Center for Nanophase Materials Science, Oak Ridge National Laboratory, Oak Ridge, Tennessee 37831 USA}
\author{Liangbo Liang} \affiliation{Center for Nanophase Materials Science, Oak Ridge National Laboratory, Oak Ridge, Tennessee 37831 USA}
\author{Santosh KC} \affiliation{Materials Science and Technology Division, Oak Ridge National Laboratory, Oak Ridge, Tennessee 37831 USA}
\author{Ercan Cakmak} \affiliation{Materials Science and Technology Division, Oak Ridge National Laboratory, Oak Ridge, Tennessee 37831 USA}
\author{Stuart Calder} \affiliation{Quantum Condensed Matter Division, Oak Ridge National Laboratory, Oak Ridge, Tennessee 37831 USA}
\author{Brian C. Sales} \affiliation{Materials Science and Technology Division, Oak Ridge National Laboratory, Oak Ridge, Tennessee 37831 USA}

\begin{abstract}
The crystallographic and magnetic properties of the cleavable $4d^3$ transition metal compound $\alpha$-\ce{MoCl3} are reported, with a focus on the behavior above room temperature. Crystals were grown by chemical vapor transport and characterized using temperature dependent x-ray diffraction, Raman spectroscopy, and magnetization measurements. A structural phase transition occurs near 585\,K, at which the Mo$-$Mo dimers present at room temperature are broken. A nearly regular honeycomb net of Mo is observed above the transition, and an optical phonon associated with the dimerization instability is identified in the Raman data and in first principles calculations. The crystals are diamagnetic at room temperature in the dimerized state, and the magnetic susceptibility increases sharply at the structural transition. Moderately strong paramagnetism in the high temperature structure indicates the presence of local moments on Mo. This is consistent with results of spin-polarized density functional theory calculations using the low and high temperature structures. Above the magneto-structural phase transition the magnetic susceptibility continues to increase gradually up to the maximum measurement temperature of 780\,K, with a temperature dependence that suggests two-dimensional antiferromagnetic correlations.

\end{abstract}

\maketitle

\section{Introduction}

Transition metal halide compounds adopt a variety of low dimensional crystal structures \cite{Lin-1993}. Layered compounds comprising 2 dimensional sheets of composition $MX_2$ or $MX_3$ (\textit{M}\,=\,transition metal, \textit{X}\,=\,halogen) van der Waals bonded to one another exhibit a range of interesting behaviors associated with their weak interlayer chemical and magnetic interactions \cite{McGuire-2017}. The \textit{M} cations form a triangular net in the dihalides and a honeycomb net in the trihalides. While these materials have been studied as examples of low dimensional magnetism and magnetic frustration for many years, recent work on layered $MX_2$ or $MX_3$ compounds is driven by interest in multiferroicity, quantum spin liquids, van der Waals heterostructures, and monolayer magnets \cite{Tokunaga-2011, Plumb-2014, Abdul-Wasey-2013, McGuire-2015}. The development of transition metal halides as cleavable materials in which long range magnetic order persists down to the limit of a single monolayer and for exchange coupling with other materials in small devices has focused mainly on the trihalides of the 3\textit{d} metals \cite{McGuire-2015, Zhang-2015, Liu-2016, Zhou-2016, He-2016}. Both ferromagnetism in monolayer specimens and strong exchange field effects in heterostructures have recently been observed with CrI$_3$ \cite{Huang-2017, Zhong-2017}. Moving to the 4\textit{d} metals increases the importance of spin-orbit coupling, and gives rise to the spin-liquid behavior associated with Kitaev interactions in RuCl$_3$ \cite{Kitaev-2006, Plumb-2014, Banerjee-2016}. Thus, there is interest in exploring magnetism in other layered 4\textit{d} and 5\textit{d} transition metal halides. Rh$X_3$ and Ir$X_3$ are isostructural to RuCl$_3$, but are expected to be non-magnetic due to their 4$d^6$ and 5$d^6$ electronic configurations and relatively strong crystal field splitting \cite{Guthrie-1931, Brodersen-1968}. Halides of the earlier heavy transition metals often form zero or 1 dimensional structures, instead of the layered structures of interest here \cite{Lin-1993}. One exception to this is \ce{MoCl3}.

\begin{figure}
\begin{center}
\includegraphics[width=3.25in]{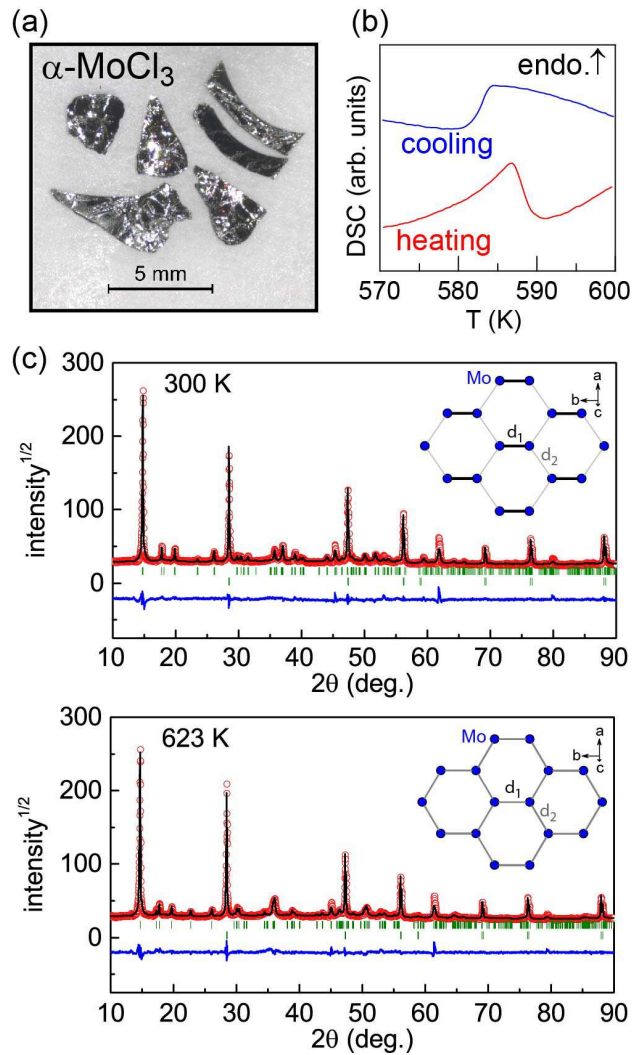}
\caption{\label{fig:diffraction}
(a) Crystals of $\alpha$-\ce{MoCl3} grown by vapor transport. (b) DSC thermal signatures of the reversible phase transition near 585\,K. (c) Rietveld fits to x-ray diffraction data collected at 300 and 623\,K with honeycomb Mo nets shown in the insets. The upper set of ticks denote reflections from AlCl$_3$-type \ce{MoCl3} and the lower set correspond to the silicon powder that was mixed with the sample.
}
\end{center}
\end{figure}

\begin{figure*}
\begin{center}
\includegraphics[width=5.25in]{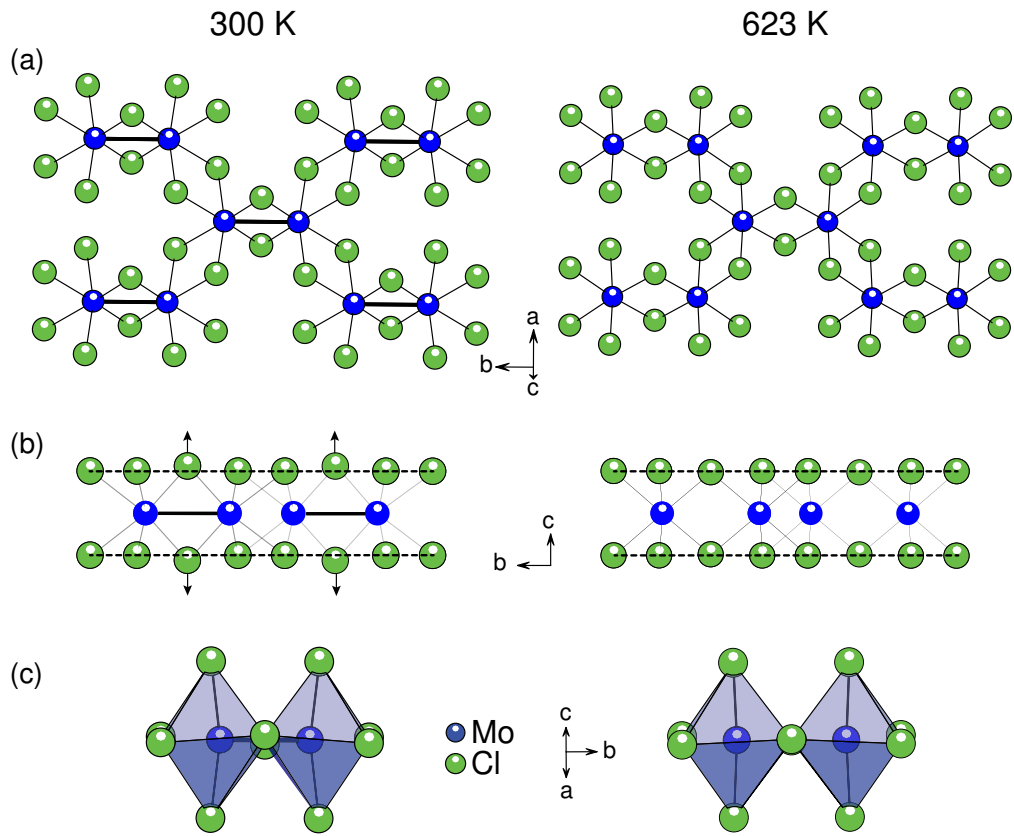}
\caption{\label{fig:struct}
The crystal structures of $\alpha$-MoCl$_3$ determined at 300 and 623\,K. (a) A single layer viewed along the stacking direction. The dimers present in the distorted honeycomb net of Mo at 300\,K are indicated by the thick black lines. (b) A single layer viewed normal to the stacking direction. Dimers are indicated by thick black lines, and dashed lines are drawn through the Cl layers to highlight the buckling out of Cl atoms between the dimerized Mo atoms as indicated by the arrows. (c) Coordination polyhedra around a single Mo-Mo pair, showing the off-centering of the Mo atoms at 300\,K and the associated bending of the apical Cl-Mo-Cl chain and slight twisting distortion of the octahedra.
}
\end{center}
\end{figure*}

$\alpha$-\ce{MoCl3} adopts the AlCl$_3$ structure type, the layered structure observed in RuCl$_3$, CrI$_3$, and several other layered trihalides. In this monoclinic structure (space group $C2/m$) the transition metal layer is not constrained by symmetry to be an ideal honeycomb net. The \textit{M} atoms reside at position (0,\,\textit{y},\,0), and the value of the y-coordinate and ratio of the \textit{a} and \textit{b} lattice parameters determine the shape of the net. Each \textit{M} has three neighbors and there are two unique $M-M$ distances within the layers. In many cases the nets are very nearly ideal, as measured by the ratio of the two in-plane $M-M$ nearest-neighbor distances $d_1$ and $d_2$. The approximate honeycomb nets of Mo with these distances labeled are shown in Figures \ref{fig:diffraction} and \ref{fig:struct}. Reported values for this ratio include $d_1/d_2$\,=\,1.000 for CrCl$_3$ \cite{Morosin-1964}, 0.998 for CrI$_3$ \cite{McGuire-2015}, and 0.999 \cite{Cao-2016} and 0.990 \cite{Johnson-2015} for RuCl$_3$. In $\alpha$-\ce{MoCl3}, however, this distortion of the honeycomb net is much larger, with reported $M-M$ in-plane distance ratios of 0.745 \cite{Schafer-1967} and 0.743 \cite{Hillebrecht-1997}. In fact, the honeycomb net in $\alpha$-\ce{MoCl3} is broken into a triangular net of Mo dimers. The intradimer distance of 2.76\,{\AA} is only slightly larger than the nearest neighbor distance in elemental Mo (2.72\,{\AA}), indicating significant Mo$-$Mo bonding. Dimer formation within the \ce{AlCl3} structure type does not change the symmetry or size of the unit cell, so it is not expected to be driven by a degeneracy-lowering Peierls distortion, but rather by Mo$-$Mo chemical bonding forming an almost molecular structure within the \ce{MoCl3} layers \cite{Lin-1993}. This type of direct cation-cation bonding can be expected for transition metals with less than half-filled d shells \cite{Goodenough-1960}. A similar dimerization is in seen other early transition metal trihalides, including TcCl$_3$ at room temperature \cite{Poineau-2012}, and TiCl$_3$ below about 220\,K \cite{Ogawa-1960}, which are discussed in more detail below.

Few reports of the physical properties of \ce{MoCl3} were found in the literature. X-ray and microscopy studies of \ce{MoCl3} have identified two layered polymorphs, $\alpha$ and $\beta$ \cite{Schafer-1967, Hillebrecht-1997}. The former adopts the AlCl$_3$ structure type described above, the latter has similar layers with a different stacking sequence reported originally in space group $C2/c$ \cite{Schafer-1967} and then in space group $C2/m$ with a significant amount of disorder, likely associated with stacking faults \cite{Hillebrecht-1997}. Magnetization measurements show $\alpha$-\ce{MoCl3} is weakly paramagnetic or diamagnetic at room temperature, depending on the precise stoichiometry, and the small magnetic moment is attributed to the structural dimerization \cite{Schafer-1967}. Heat capacity was reported from 5 to 350\,K and found no thermal anomalies \cite{Kiwia-1975}. Hillebrecht et al. noted a phase transition near 583\,K in differential scanning calorimetry measurements, and attributed changes in the x-ray diffraction pattern at high temperatures to the breaking of the Mo$-$Mo dimers \cite{Hillebrecht-1997}, but no structural information for the high temperature phase was given and magnetic consequences were not addressed.


Here we examine the magnetic and structural properties of $\alpha$-\ce{MoCl3} crystals, focusing on behavior observed above room temperature. We employ X-ray diffraction to study the crystallographic phase transition, and report lattice parameters and interatomic distances as a function of temperature up to 623\,K, identifying and refining the high temperature structure. We also present the evolution of the measured Raman spectra. Our high temperature magnetic susceptibility data reveal a strong anomaly near the structural transition at 585\,K. Upon warming, the magnetic susceptibility increases dramatically, and the material transforms from a diamagnet near room temperature to a relatively strong paramagnet at high temperature. We also observe the transition in differential scanning calorimetry measurements, and both measurements show the thermal hysteresis associated with the phase transition is limited to a few degrees or less.  We identify the magnetic response as the unlocking of paramagnetic \textit{S}\,=\,$\frac{3}{2}$ moments of individual Mo$^{3+}$ cations as the dimer states are destroyed. Interestingly, the magnetic susceptibility does not exhibit Curie-Weiss like behavior above the transition. Measurements up to 780\,K show a gradual increase in susceptibility with a negative curvature that suggests a broad maximum may be expected at higher temperature. This is behavior typical of layered magnetic materials with strong antiferromagnetic in-plane magnetic exchange and weak interplanar interactions. If this scenario is correct, a maximum in susceptibility above 800\,K would indicate very strong antiferromagnetic interactions in the high temperature phase of \ce{MoCl3}, which is also indicated by results of our density functional theory calculations.

\section{Results and discussion}

Details of the synthesis and characterization techniques can be found below in the Methods Section. Crystals of \ce{MoCl3} grown for this study using a vertical vapor transport technique are shown in Figure \ref{fig:diffraction}a. Reactions typically produce intergrown platelike crystals that can be easily separated from one another and cleaved into thin samples. Differential scanning calorimetry curves (Figure \ref{fig:diffraction}b) show a reversible phase transition near 585\,K, consistent with Ref. \citenum{Hillebrecht-1997}.

Results from powder x-ray diffraction measurements carried out at 300 and 623\,K are summarized in Figure \ref{fig:diffraction}c, Figure \ref{fig:struct}, and Table \ref{tab:structure}. At both temperatures the data are well described by the monoclinic AlCl$_3$ structure type with [00l] texture, confirming that the crystals are single phase $\alpha$-\ce{MoCl3}. Refinement without including the [00l] texture did not significantly change the values of the key structural parameters. For example, y$_{\rm{Mo}}$ changed by only 2\%, to 0.145. The difference between the high and low temperature structure is primarily in the Mo$-$Mo distances in the honeycomb nets, as proposed by Hillebrecht et al. \cite{Hillebrecht-1997}. The net is broken by strong dimerization at 300\,K, with the intradimer distance almost 1\,{\AA} shorter than the interdimer distance. At 623\,K, the distances within the Mo net vary by less than 0.15\,{\AA}, with a $d_1/d_2$ ratio of 0.959. Though this deviation is small, it does indicate some distortion of the average structure at high temperature away from an ideal honeycomb net, either in the form of a small static cooperative distortion, a residual small fraction of directionally coherent dimers, or peraps a more complex disordered structure like that identified in \ce{Li2RuO3} \cite{Kimber-2014}and discussed in more detail below.

The low and high temperature crystal structures of MoCl$_3$ are shown in Figure \ref{fig:struct}. The distortion of the Mo net results in a notable distortion of the anion sublattice as well, as seen in the buckling of the Cl layers, highlighted in Figure \ref{fig:struct}b. In the high temperature, undimerized state, the Cl layers are flatter and the distribution of Mo$-$Cl distances is narrower (Table \ref{tab:structure}). Figure \ref{fig:struct}c shows the coordination around a pair of Mo atoms that make up the dimers in the low temperature structure. The strong off-centering of the Mo ions and associated twisting distortion of the Cl octahedra are both significantly relaxed at 623\,K.

\begin{table}
\caption{\label{tab:structure} Results of Rietveld refinement of powder x-ray diffraction data from $\alpha$-\ce{MoCl3} collected at the indicated temperatures. The structure is described by space group $C2/m$ at both temperatures with Mo at (0,y,0), Cl1 at (x,y,z), and Cl2 at (x,0,z).}
\begin{tabular}{lcc}					
\hline
T (K)	&	300	&	623	\\
a ({\AA})	&	6.1120(10)	&	6.0353(10)	\\
b ({\AA})	&	9.7816(18)	&	10.2842(16)	\\
c ({\AA})	&	6.3146(4)	&	6.3592(4)	\\
$\beta$ (deg.)	&	108.16(1)	&	108.32(1)	\\
y$_{\rm{Mo}}$	&	0.1420(8)	&	0.1634(7)	\\
x$_{\rm{Cl}1}$	&	0.250(2)	&	0.268(3)	\\
y$_{\rm{Cl}1}$	&	 0.3291(15)	&	0.3267(13)	\\
z$_{\rm{Cl}1}$	&	0.2320(9)	&	0.2418(10)	\\
x$_{\rm{Cl}2}$	&	0.230(3)	&	0.231(4)	\\
z$_{\rm{Cl}2}$	&	0.2600(14)	&	0.2352(17)	\\
Mo-Mo $d_1$, $d_2$ ({\AA})	&	2.778(11), 3.775(6)	&	3.361(10), 3.504(5)	\\
Mo-Cl ({\AA})	&	2.272(10)$-$2.537(13)	&	 2.385(15)$-$2.497(13)	\\
Rp, Rwp	&	6.6, 9.2	&	5.8, 8.3	\\
\hline				
\end{tabular}	
\end{table}

\begin{figure}
\begin{center}
\includegraphics[width=3.0in]{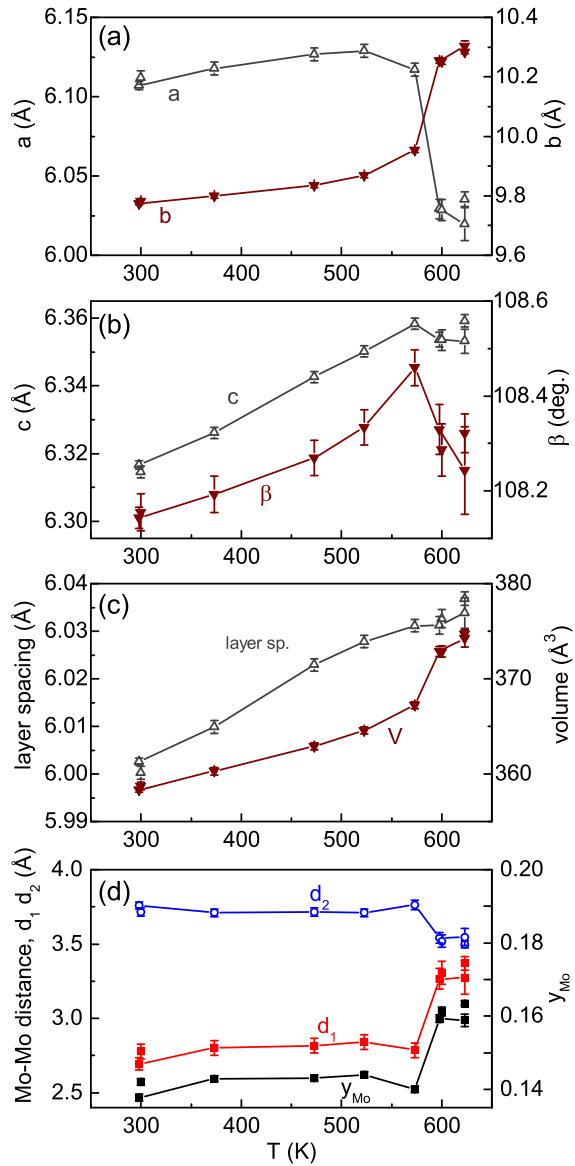}
\caption{\label{fig:lattice}
Temperature dependence of the structural parameters determined from diffraction measurements on $\alpha$-\ce{MoCl3}. (a) The \textit{a} and \textit{b} lattice parameters. (b) The \textit{c} lattice parameter and monoclinic angle $\beta$. (c) The layer spacing corresponding to the perpendicular distance between midpoints of neighboring layers and the unit cell volume. (d) The nearest neighbor Mo$-$Mo distances $d_1$ and $d_2$ as defined in Fig. \ref{fig:diffraction}c, and the \textit{y}-coordinate of Mo. Results from a second sample at 300, 600, and 623\,K are also shown.
}
\end{center}
\end{figure}

The temperature dependence of crystallographic parameters between room temperature and 623\,K are shown in Figure \ref{fig:lattice}. The crystallographic phase transition is apparent in all of the lattice parameters (Fig. \ref{fig:lattice}a,b). As the dimers break upon heating, \textit{b} increases and \textit{a} decreases. This is consistent with the dimers being oriented along the \textit{b}-axis. The parameters that relate to the layer stacking, \textit{c} and $\beta$, also respond to the structural transition; however, the layer spacing (\textit{c}\,sin$\beta$) varies fairly smoothly (Fig. \ref{fig:lattice}c).

The unit cell volume, shown in Figure \ref{fig:lattice}c, increases abruptly by about 1.6\% at the transition, consistent with additional Mo$-$Mo bonding in the low temperature state. The Mo$-$Mo interatomic distances $d_1$ and $d_2$ are shown in Figure \ref{fig:lattice}d. These distances are defined in Figure \ref{fig:diffraction}c, and at room temperature $d_1$ represents the intradimer distance and $d_2$ the shortest interdimer distance. The temperature dependence of these Mo-Mo distances are dominated by the internal coordinate \textit{y} of the Mo atom. The values of $y_{\rm{Mo}}$ remain nearly temperature independent up to 573\,K, and then increases by about 16\% between 573 and 600\,K (Fig. \ref{fig:lattice}d).

\begin{figure}
\begin{center}
\includegraphics[width=3.0in]{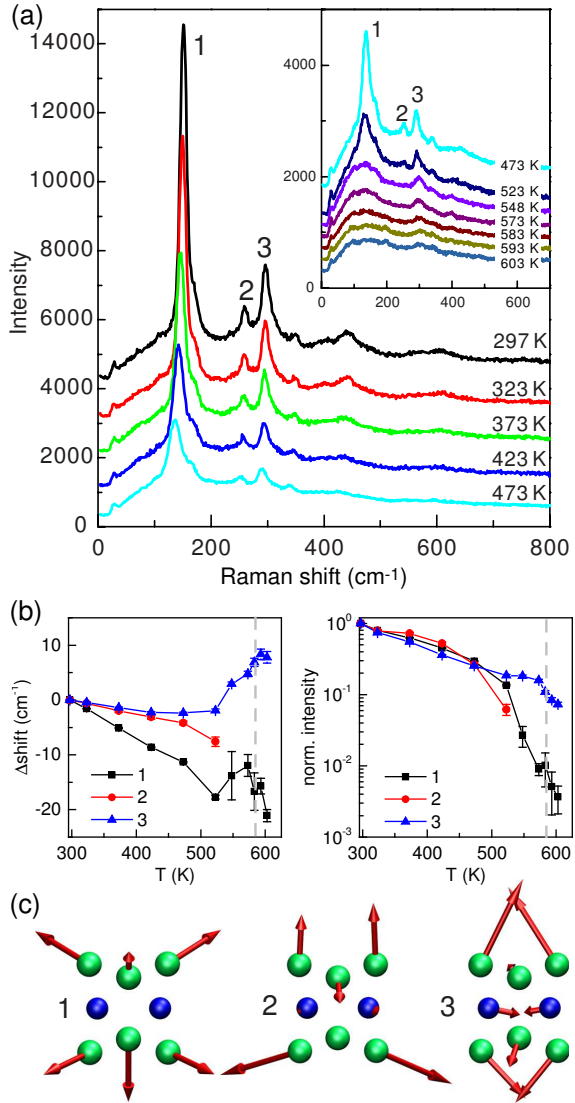}
\caption{\label{fig:Raman}
Results of Raman spectroscopy measurements on $\alpha$-\ce{MoCl3}. (a) Spectra collected at the temperatures indicated on the plot, offset vertically. (b) Temperature dependence of the Raman shifts and intensities relative to 297\,K values for modes 1, 2, and 3 labeled on panel (a). (c) Atomic displacements corresponding to modes 1, 2, and 3.
}
\end{center}
\end{figure}

The structural transition in MoCl$_3$ was also tracked with Raman spectroscopy. Spectra collected at temperatures from 297 to 603\,K are shown in Figure \ref{fig:Raman}a. Three main peaks labeled 1, 2, and 3 are seen at 148\,cm$^{-1}$, 256\,cm$^{-1}$, and 294\,cm$^{-1}$, respectively, at room temperature. Several observations can be made using the data in Figure \ref{fig:Raman}. The temperature dependence of the positions of the labeled peaks behave normally up to 473\,K. The Raman shifts decrease as the lattice softens upon heating, and the  decrease in intensity with temperature is consistent with the Bose factor. Upon further heating several anomalous behaviors are observed, which may be associated with the crystallographic phase transition. Modes 1 and 2 show a rapid drop in intensity and mode 3 hardens significantly (Figure \ref{fig:Raman}b). These behaviors onset near 550\,K, somewhat lower than the phase transition temperature. Upon cooling back to room temperature, relatively sharp spectra were recovered, but with intensity reduced to about 1/3 of that seen before heating. This is attributed to a buckling of the crystal surface that occurs on cycling through the structural phase transition.

The spectra in Figure \ref{fig:Raman} are qualitatively similar to spectra reported recently for RuCl$_3$ \cite{Glamazda-2017}. Upon heating through the rhombohedral to monoclinic structural transition in that material and in isostructural \ce{CrCl3}, several modes were observed to harden by 1$-$3\,cm$^{-1}$. This is small compared to the change in mode 3 for \ce{MoCl3} (Fig. \ref{fig:Raman}), but the phase transitions in these materials are quite different. In \ce{RuCl3} the transition is associated with a change in the stacking sequence, which is expected to influence mainly the weak van der Waals interactions between layers, while in \ce{MoCl3} the transition affects the stronger in-plane bonding.

First principles phonon calculations to examine zone center phonon energies and vibrational displacements were performed to compare with the experimental data. Using the structure determined from minimization of the energy and forces within the local density approximation (LDA) ($d_1$\,=\,2.5568\,{\AA} and $d_2$\,=\,3.6253\,{\AA}) Raman active modes were identified. The vibrational energies are expected to be overestimated due to overbinding associated with the LDA and lack of thermal expansion in the calculated values, which tends to decrease mode frequencies. The calculations show Raman active modes at 202\,cm$^{-1}$, 289\,cm$^{-1}$, and 335\,cm$^{-1}$ that are identified with modes 1, 2, and 3, respectively. There is a calculated mode at 220\,cm$^{-1}$ that may also contribute to the peak labeled mode 1. The LDA structure was then distorted by increasing the intradimer Mo-Mo distance, to simulate the experimentally observed structural phase transition. After relaxing the Cl positions with the unit cell and Mo positions fixed, phonon calculations were repeated. With the Mo-Mo distances changed to $d_1$\,=\,2.757\,{\AA} and $d_2$\,=\,3.510\,{\AA}, Mode 3 was observed to harden, moving to 364\,cm$^{-1}$, while other modes did not, consistent with the experimental observations. This effect was most clear for small distortions, as it was difficult to follow individual modes when the distortion was large. The calculated atomic displacements associated with modes 1, 2, and 3 are shown in Figure \ref{fig:Raman}c. Interestingly, of these vibration only mode 3, which behaves anomalously near the phase transition, has a significant Mo-Mo dimerization character.

\begin{figure}
\begin{center}
\includegraphics[width=3.0in]{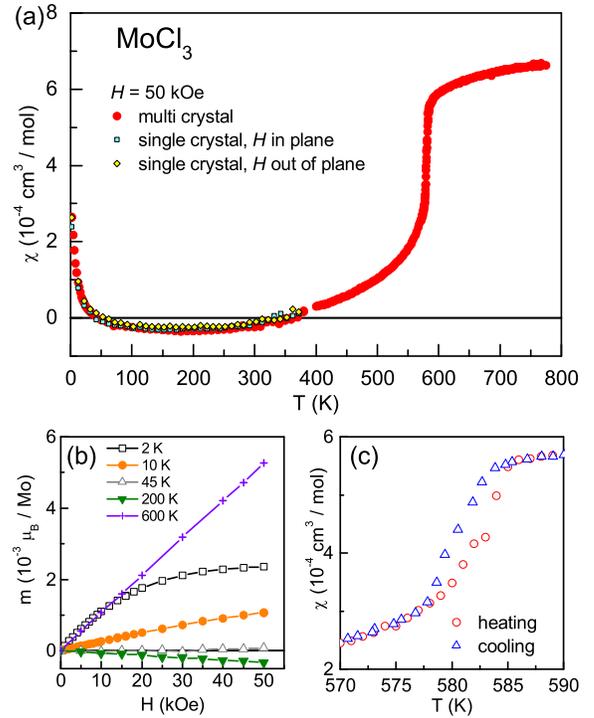}
\caption{\label{fig:mag}
The crystallographic phase transition in $\alpha$-\ce{MoCl3} probed by magnetization measurements. (a) Magnetic susceptibility $\chi = M/H$ of \ce{MoCl3} crystals measured in an applied field of 50\,kOe. (b) Isothermal magnetization curves at selected temperatures. (c) Susceptibility determined upon heating and cooling through the phase transition region.
}
\end{center}
\end{figure}

As noted above, the magnetic susceptibility ($\chi = M/H$) of \ce{MoCl3} is known to be close to zero at room temperature, and this is attributed to the formation of an $S = 0$ state associated with the strong Mo$-$Mo dimerization. Thus a strong response of the magnetic properties to the structural phase transition may be expected. Figure \ref{fig:mag}a shows the magnetic susceptibility of $\alpha$-\ce{MoCl3} crystals between 2 and 780\,K, with isothermal magnetization curves at selected temperatures shown in Figure \ref{fig:mag}b. The data collected at and below room temperature will be discussed first, and then behavior near the structural phase transition will be addressed.

A diamagnetic signal is observed at room temperature and down to about 45\,K. The low temperature behavior is dominated by a small Curie tail attributed to defects or impurities contributing local paramagnetic moments. Molybdenum vacancies would provide one possible source for such moments, as they would leave undimerized Mo atoms behind. If this is the case, the size of the low temperature tail in Figure \ref{fig:mag}a would indicate about 0.003 Mo-vacancies per formula unit. A Brillouin function fit to the 2\,K M vs H data suggests about 0.001 spin-$\frac{3}{2}$ per formula unit. Similarly, Cl vacancies could be a source of dilute Mo$^{2+}$ (\textit{S}\,=\,1) local moments. Nearly identical magnetic behaviors near and below room temperature (Figure \ref{fig:mag}a) are seen in a multi-crystal sample grown using conventional vapor transport and a single crystal grown under different conditions using a vertical vapor transport configuration (see Methods). No magnetic anisotropy is detected in the single crystal, which rules out long range antiferromagnetic order as the origin of the low magnetic susceptibility, since such a state would have associated magnetic anisotropy. This conclusion is consistent with preliminary neutron powder diffraction measurements conducted at room temperature, which showed only reflections indexed by the crystallographic unit cell.

Upon heating above room temperature, the magnetic susceptibility increases gradually as the structural phase transition is approached. At the transition, a sharp increase in magnetization is observed. This sharp increase occurs between about 580 and 585\,K. Comparison of data collected upon heating and cooling through the transition shows that the thermal hysteresis is small, and similar to the small hysteresis observed in the DSC data (Fig. \ref{fig:diffraction}b). The abruptness of the transition and the small thermal hysteresis suggest that the dimerization transition may be first order.

The behavior of the magnetic susceptibility above the structural phase transition in \ce{MoCl3} is particularly interesting. As seen in Figure \ref{fig:mag}a, the susceptibility continues to increase upon heating up to the maximum measurement temperature of 780\,K. That is, Curie Weiss paramagnetism is not observed in the high-temperature, dimer-free structure even up to temperatures 30\% larger than the dimerization temperature. This raises the question of whether the Mo 4$d$ electrons are localized or not above 585\,K. The magnitude of the susceptibility at 600\,K is $6\times10^{-4}$\,cm$^3$/mol. This is an order of magnitude larger than typical values of the Pauli paramagnetic susceptibility of metals. It is therefore unlikely that the high temperature behavior is associated with delocalized Mo 4$d$ electrons, and the magnetic susceptibility is expected to arise from localized moments.

The measured high temperature susceptibility is also significantly smaller than the estimate for non-interacting local moments of spin-$\frac{3}{2}$, which is $3\times10^{-3}$\,cm$^3$/mol at 600\,K (the orbital moment is expected to be quenched, with each $t_{\textrm{2g}}$ orbital singly occupied). This suggests that there are strong antiferromagnetic correlations in the high temperature phase that suppress the susceptibility and produce the unusual temperature dependence seen above 585\,K in Figure \ref{fig:mag}a. Above this temperature, the susceptibility shows negative curvature and one can speculate that it may go through a broad maximum at some temperature greater than 800\,K. Qualitatively, this is the type of temperature dependence often observed in (quasi-) 2D materials with strong antiferromagnetic interactions \cite{LayeredTM}. This would indicate that the magnetic interactions in \ce{MoCl3} are quite strong. For $S = \frac{3}{2}$ on a honeycomb lattice, the susceptibility maximum is expected to occur at a temperature close to $1.5 |J| S(S+1)/k_B$ \cite{Navarro-1990}. Supposing that the maximum occurs near or above 800\,K would then give $|J|/k_B \geq 140$\,K. Alternatively, the observed behavior could indicate long range magnetic order in the high temperature state. Although this may seem unlikely, the $4d^3$ electronic configuration is associated with high magnetic ordering temperatures. Examples include SrTcO$_3$ \cite{Rodriguez-2011} with a N\'{e}el temperature near 1000\,K and SrRu$_2$O$_6$ \cite{Tian-2015, Hiley-2015} with a N\'{e}el temperature of 565\,K. As described further below, calculations indeed predict strong antiferromagnetic interactions in \ce{MoCl3}.

Similar dimerizations have been noted in other honeycomb lattice materials. TiCl$_3$ has an ideal honeycomb net of Ti$^{3+}$ at room temperature and a sharp drop in magnetic susceptibility is seen upon cooling below about 220\,K \cite{Klemm-1947, Ogawa-1960}. Though originally expected to be associated with magnetic order, a lattice anomaly was observed at the same temperature and it was speculated that the behavior may be related to metal-metal bonding \cite{Ogawa-1960, Goodenough-1960}. The transition is apparent in other physical properties as well, including electrical resistivity. The in-plane resistivity decreases by about a factor of ten upon warming through the transition region, but TiCl$_3$ is semiconducting over the entire temperature range \cite{Cavallone-1970}. A detailed structural investigation revealed a distortion of the honeycomb net at the transition, in which dimers are formed resulting in in-plane Ti-Ti distances of 3.37 and 3.58\,{\AA} \cite{Troyanov-1991}. The strength of the dimerization, as measured by the difference in $M-M$ distances, is weaker in TiCl$_3$ than in \ce{MoCl3}, presumably due to the number of $d$ electrons involved and the greater extent of the 4$d$ orbitals of Mo. Strong dimerization is also seen in TcCl$_3$ at room temperature \cite{Poineau-2012, Poineau-2013}. In analogy with TiCl$_3$ and \ce{MoCl3}, a dimer-breaking transition to a more strongly magnetic state may be expected at high temperature in TcCl$_3$ as well.

The honeycomb ruthenate Li$_2$RuO$_3$ is known to undergo a dimerization transition upon cooling below about 540\,K \cite{Miura-2007}. The magnetic behavior of this material is very similar to that shown for MoCl$_3$ in Figure \ref{fig:mag} \cite{Miura-2007, Mehlawat-2017}. The average structure of Li$_2$RuO$_3$ determined by diffraction methods above the transition contains a near-perfect honeycomb Ru net.  Kimber et al. used first principles calculations and x-ray pair distribution function analysis to demonstrate that the local structure at high temperature is in fact a disordered dimer or valence bond liquid state \cite{Kimber-2014}, in which the dimers persists above the transition but the long range order is destroyed by thermal fluctuations. Such measurements are desirable to investigate the possibility of similar local structures in  MoCl$_3$ (and TiCl$_3$). As noted above, a small but measurable distortion in the average structure is indicated by x-ray diffraction above the transition in MoCl$_3$ (Fig. \ref{fig:lattice}d). Thus, if a high temperature disordered dimer model is correct for MoCl$_3$, there must be more spatial coherence than observed in \ce{Li2RuO3}.

\begin{table}
\caption{\label{tab:dft} Results of magnetic vdW-DF-C09 calculations for \ce{MoCl3} in the low-temperature (LT) and high-temperature (HT) structures with relaxed and experimental atomic positions. Energies are given relative to the non-magnetic state.
}
\setlength{\tabcolsep}{2mm}
\begin{tabular}{cccc}					
\hline
Structure   &	Magnetic order	               &	Moment	       &	$\Delta$E	\\
	        &	intraplanar - interplanar	   &	($\mu_B$ / Mo)	       &	(eV / Mo)	\\
\hline
LT relaxed	 &	FM-FM	 &	1.1	 &	-0.019	\\
LT relaxed	 &	AFM-FM	 &	1.6	 &	-0.112	\\
LT exp	 &	FM-FM	 &	2.1	 &	0.151	\\
LT exp	 &	AFM-FM	 &	1.8	 &	-0.121	\\
HT exp	 &	FM-FM	 &	3.1	 &	-0.189	\\
HT exp	 &	AFM-FM	 &	2.8	 &	-0.336	\\
HT exp	 &	FM-AFM	 &	3.0	 &	-0.195	\\
HT exp	 &	AFM-AFM	 &	2.8	 &	-0.338	\\

\hline				
\end{tabular}	
\end{table}

To further investigate the magnetic properties of \ce{MoCl3}, first principles calculations using spin-polarized vdW-DF-C09 were performed. Three different crystal structures were used: the low and high temperature structures with the unit cells and atom positions fixed at the experimental values given in Table \ref{tab:structure}, and a structure in which the atomic coordinates were allowed to relax starting with the experimental low temperature structure. When the atomic positions in the high temperature structure were not held fixed, the structure relaxed into a dimerized state in a non-spin-polarized calculation, indicating that the dimerization is driven primarily by Mo$-$Mo chemical bonding interactions and not magnetism.

Several magnetic configurations were tested in the calculations. The results are shown in Table \ref{tab:dft}, where calculated magnetic moment on each Mo ion and energies relative to the energy of the non-magnetic state are listed. The magnetic structures are labeled by intraplanar and interplanar orders (FM for ferromagnetic, AFM for antiferromagnetic). The intraplanar AFM structure used is the simple N\'{e}el state with all nearest neighbors antialigned.

It is interesting to note how the size of the moment varies among the different structures considered in the calculations. In the low temperature structures the moment is strongly reduced below the value of 3\,$\mu_B$ expected for $4d^3$ trivalent Mo. This is consistent with the formation of chemically bonded Mo$-$Mo pairs. The Mo magnetic moment increases as the dimers are broken, and in the high temperature structure a value close to the full 3\,$\mu_B$ per Mo is observed.

Comparison of the energies in Table \ref{tab:dft} shows that the intraplanar AFM configurations are the most stable. Since a local magnetic moment is expected to be present in the undimerized state, it is interesting to estimate magnetic exchange interaction parameters from the computed energies in the high temperature structure. This is done using a Heisenberg model Hamiltonian $H=- \sum J_{ij} S_i \cdot S_j$ with the sum over Mo-Mo pairs that interact with exchange coupling $J_{ij}$ and with spin $S = \frac{3}{2}$. If only a single in-plane $J$ is considered between a Mo site and its three nearest neighbors, it can be evaluated from the energy difference between the FM-FM and the AFM-FM structures by $J = \frac{E_{AFM-FM}-E_{FM-FM}}{3S^2}$, giving $J_1/k_B$\,=\,-255\,K (negative indicating AFM). However, the small energy difference between the FM and AFM stacked structures for both types of intraplanar order shows that significant interplanar coupling is present. The two closest interlayer Mo-Mo distances are 6.36 and 6.38\,{\AA}, one of each in the plane above, and one of each in the plane below. Since these differ by only 0.02\,{\AA} they will both be considered, and since the energy of the AFM-FM and AFM-AFM structures are nearly the same the difference in exchange between these two interplanar interactions is near the limit of the calculation precision. Assuming a single intraplanar exchange with $z_{\mathrm{intra}}$\,=\,3 neighbors and a single interplanar exchange with $z_{\mathrm{inter}}$\,=\,4, the energies in Table \ref{tab:dft} give estimates of  $J_{\mathrm{intra}}/k_B$\,=\,-250\,K and $J_{\mathrm{inter}}/k_B$\,=\,-8\,K, or $J_{\mathrm{intra}}z_{\mathrm{intra}}/k_B$\,=\,-750\,K and $J_{\mathrm{inter}}z_{\mathrm{inter}}/k_B$\,=\,-32\,K.

These results can be compared for the analogous 3$d$ transition metal compounds CrCl$_3$ and CrBr$_3$ for which experimentally determined exchange strengths have been reported. The values for CrCl$_3$ are $J_{\mathrm{intra}}z_{\mathrm{intra}}/k_B$\,=\,15.75\,K and $J_{\mathrm{inter}}z_{\mathrm{inter}}/k_B$\,=\,-0.037\,K \cite{Narath-1965}, and the values for CrBr$_3$ $J_{\mathrm{intra}}z_{\mathrm{intra}}/k_B$\,=\,24.75\,K and $J_{\mathrm{inter}}z_{\mathrm{inter}}/k_B$\,=\,0.99\,K \cite{Davis-1964}. The in-plane nearest neighbor J has also been computed for these two compounds using methods similar to those used here, and values of $J_{\mathrm{intra}}/k_B$\,=\,22\,K for CrCl$_3$ and 29\,K for CrBr$_3$ are reported \cite{Zhang-2015}. The relatively weak interlayer coupling in these materials is expected to suppress the development of long range magnetic order. CrCl$_3$ and CrBr$_3$ do order magnetically below about 17\,K and 37\,K, respectively \cite{Tsubokawa-1960-CrBr3, Cable-1961-CrCl3}. The products $Jz$ both in and out of plane for MoCl$_3$ are at least 30 times larger than the experimental values reported for these Cr compounds, and the in-plane interactions are about a factor of 10 larger in MoCl$_3$  than the calculated values for CrCl$_3$ and CrBr$_3$. This suggests that a significantly higher magnetic ordering temperature might be possible in MoCl$_3$ if it were not interrupted upon cooling by the moment quenching dimerization.

\section{Summary and conclusions}

The dimer-breaking, magneto-structural phase transition in \ce{MoCl3} is observed upon heating near 585\,K via x-ray diffraction, calorimetry, Raman spectroscopy, and magnetic susceptibility measurements. The room temperature \ce{AlCl3} structure type is retained at high temperature. Although the symmetry is unchanged, the extreme distortion of the honeycomb net present at room temperature is relieved as the Mo-Mo dimers break and a nearly regular honeycomb net is observed at 623\,K, with only about a 5\% difference between Mo-Mo in-plane distances above the transition. Anomalous behavior of a Mo-dimer phonon mode associated with the phase transition is identified by Raman spectroscopy. Magnetism is quenched by the strong Mo-Mo interactions at room temperature but large local magnetic moments emerge in the high temperature state. Although the data do not extend to high enough temperature to draw definitive conclusions about the nature of the high temperature magnetic behavior, the results suggest the presence of strong antiferromagnetic coupling between Mo S\,=\,$\frac{3}{2}$ moments. First principles calculations also support strong antiferromagnetic exchange within the honeycomb Mo nets, with significantly weaker interlayer interactions. All of the magnetic interactions appear to be stronger by at least an order of magnitude than in the analogous 3$d$ transition metal compounds CrCl$_3$ and CrBr$_3$, suggesting that higher transition temperatures may be possible in MoCl$_3$, if the dimerization transition were avoided. The calculations give a magnetic moment consistent with spin-$\frac{3}{2}$ in the high temperature structure and a strong reduction of the moment in the dimerized state. The behavior observed here is similar to that reported for \ce{TiCl3}, though the magnetic response is much stronger in \ce{MoCl3} and the transition temperature is much higher. A similar coupled structural and magnetic transition may be expected in \ce{TcCl3}, which is also known to have a dimer structure at room temperature. In analogy with Li$_2$RuO$_3$, the possibility of a valence bond liquid state of disordered dimers at high temperature in these materials should be investigated with local structure probes like pair distribution function analysis of scattering data. The high temperature magnetic properties of \ce{MoCl3} motivates future chemical substitution studies in this compound aimed at destabilizing the dimers and extending the magnetism to lower temperatures where magnetic order may be observed.

\section{Methods}

Crystals of \ce{MoCl3} were obtained by chemical vapor transport reactions with multiple configurations. The crystals used for most of the experimental studies reported were grown by reacting molybdenum powder (99.999\%) with molybdenum(V) chloride (99.6\%) in a 2:3 molar ratio inside evacuated and sealed silica tubes of inner diameter 16\,mm, and wall thickness 1.5\,mm. The sealed quartz tubes were first kept in a box furnace at 400$^{\circ}$C and then placed in a one zone tube furnace with the starting chemicals at the hot end with a setpoint temperature of 425$^{\circ}$C. \ce{MoCl3} crystals grew at the colder end of the tube which was at 380$^{\circ}$C. The crystal growth proceeded rapidly and all chemicals deposited at the cold end in about 8 hours. This produced clustered, thin, plate-like crystals with lateral dimensions of up to a few mm. Larger crystals like those shown in Fig. \ref{fig:diffraction}a and used for the anisotropic magnetization measurements were grown by vapor transport in a sealed silica ampoule in a vertical configuration. MoCl$_5$ and \ce{MoCl3} were loaded into a quartz crucible, which was kept near the top of a quartz ampoule of length $\approx$13\,cm by means of a smaller diameter quartz support rod. The ampoule was loaded into a Bridgman style furnace. The furnace was then heated to 425$^{\circ}$C and the sample  then lowered from a region of homogenous temperature into a cooler region at a rate of 1\,mm/h for 65h, at which point the furnace was turned off to cool naturally. Crystals were found in the top region of the ampoule within the elevated quartz crucible, and excess Mo-Cl binaries were primarily located at the bottom of the ampoule (which was the coolest spot).

X-ray diffraction measurements were conducted with a PANalytical X'Pert Pro powder diffractometer using Cu-K$_{\alpha}$ radiation. The temperature was controlled with an Anton Paar XRK 900 Reactor Chamber. The sample was prepared by gently grinding \ce{MoCl3} crystals with silicon to obtain a fine powder. This powder was then annealed in an evacuated silica ampoule at 300\,$^{\circ}$C overnight to relieve the strain induced by grinding. Data was collected from room temperature up to 623\,K under flowing ultrahigh purity (UHP) helium, and all data was analyzed by Rietveld refinement using FullProf at room temperature. Magnetization measurements were preformed with a Quantum Design MPMS SQUID magnetometer. For measurements below 400\,K a 14\,mg cluster of crystals grown by conventional vapor transport and a 8.7\,mg single crystal grown by vertical vapor transport were attached to thin silica rods with a small amount of GE varnish and placed inside a plastic drinking straw. For measurements above 400\,K several crystals with total mass 21.3\,mg were sealed inside a thin silica tube with about $\frac{1}{3}$ of an atmosphere of UHP argon. Differential scanning calorimetry data was collected with a Perkin Elmer Diamond Pyris DSC in an UHP argon atmosphere using a 7\,mg sample and a temperature ramp rate of 20\,K per minute. Powder neutron diffraction data were collected on a coarse grained polycrystalline sample comprising small crystals grown by vapor transport on the Neutron Powder Diffractometer (HB-2A) at the High Flux Isotope Reactor.

Temperature dependent Raman spectra were measured using a custom micro-Raman setup with a continuous wave solid-state laser (Excelsior, Spectra-Physics, wavelength 532 nm) as an excitation source and a 50x-long working distance microscope objective with NA (numeric aperture)\,=\,0.5 (beam spot on the samples was $\sim$1.5\,$\mu$m).  All measurements were carried out under a microscope in backscattering configuration with linear polarized excitation and unpolarized detection.  The scattered Raman light was analyzed by a spectrometer (Spectra Pro 2300i, Acton, \textit{f} \,=\,0.3\,m) that was coupled to a microscope and equipped with a 1800 groves/mm grating and a CCD camera (Pixis 256BR, Princeton Instruments). All measurements were conducted in a high temperature microscope stage (Linkam, TS 1500) under flowing Ar at atmospheric pressure.

Calculations for comparing crystallographic and magnetic structures were performed using the Quantum Espresso simulation package (v. 6.0). A planewave energy cutoff of 80 Ry was employed for the wavefunctions and 800 Ry on the charge density with a $4\times8\times4$ Monkhorst-Pack k-point mesh. Ultrasoft pseudopotentials from the Garrity-Bennet-Rabe-Vanderbilt (GBRV) database were employed with electronic configurations of Mo $4s^2$ $4p^6f$ $4d^4$ $5s^2$ and Cl $3s^2$ $3p^5$. To account for the van der Waals interactions that occur between layers, the vdW-DF non-local correlation functional \cite{Dion-2004, Thonhauser-2007} with the C09 exchange functional \cite{C09} was used. Both the high- and low-temperature structures were fixed to the experimental lattice constants and ionic positions. For the low temperature structure a full ionic relaxation at the experimental lattice constants was performed (until all the forces were less than 3$\times$10$^{-4}$\,Ry/bohr) and the structure was found not to  deviate significantly from the fixed coordinate structure. Inclusion of an on-site U (=2 eV) was not found  to significantly improve the agreement between the structural parameters and the experimental values. In PBE calculations performed with and without spin-orbit coupling the orbital moment on Mo was found to be negligible (0.01  B).  Based on these observations, no U and no spin-orbit coupling was included in the final calculations.

Zone-center vibrational frequencies were calculated from density functional perturbation theory using the Quantum Espresso package \cite{Baroni-2001, Gianozzi-2009} within the local density approximation (LDA).  Electronic structure was determined using a 4$\times$4$\times$3 k-mesh with plane-wave energy cutoff 50 Ry.  The relaxed structural parameters are:  \textit{a}\,=\,5.8404\,{\AA}, \textit{b}\,=\,9.4102\,{\AA}, \textit{c}\,=\,5.9750\,{\AA}, $\beta$\,=\,106.68\,$^{\circ}$, with Mo-Mo in plane distances of 2.5568\,{\AA} and 3.6253\,{\AA}.  These are significantly smaller than the measured values for the low temperature structure in Table \ref{tab:structure}.  This is expected as LDA calculations typically overbind atoms \cite{Haas-2009} and the effects of lattice expansion with temperature are not included. Therefore, we expect calculated vibrational frequencies to be somewhat overestimated.

\section*{Acknowledgements}
The authors acknowledge David Parker for helpful discussion. Research supported by the U. S. Department of Energy, Office of Science, Basic Energy Sciences, Materials Sciences and Engineering Division (crystal growth, crystallographic and magnetic studies).
A portion of this research at ORNL's High Flux Isotope Reactor was sponsored by the Scientific User Facilities Division, Office of Basic Energy Sciences, US Department of Energy.
The Raman spectroscopy was conducted at the Center for Nanophase Materials Sciences, which is a DOE Office of Science User Facility.
P.L.-K. acknowledges support from the Gordon and Betty Moore Foundation EPiQS Initiative Grant GBMF4416, and L.L. was supported by Eugene P. Wigner Fellowship at ORNL.
Computational resources were provided by the National Energy Research Scientific Computing Center (NERSC), which is supported by the Office of Science of the U.S. DOE under Contract No. DE-AC02-05CH11231.


%

\end{document}